\documentstyle[12pt] {article}
\newcommand{\beq}{\begin{equation}}
\newcommand{\eeq}{\end{equation}}
\newcommand{\beqa}{\begin{eqnarray}}
\newcommand{\eeqa}{\end{eqnarray}}
\newcommand{\ba}{\begin{array}}
\newcommand{\ea}{\end{array}}

\begin{document}

\begin{center}
{\large \bf Reply to a Comment on "the Role of Dimensionality in the 
Stability of a Confined Condensed Bose Gas"}
\end{center}

\vskip 0.5 truecm

\begin{center}
{\bf Luca Salasnich}
\footnote{E--Mail: salasnich@padova.infn.it}
\vskip 0.4 truecm
Dipartimento di Matematica Pura ed Applicata,\\
Universit\`a di Padova, Via Belzoni 7, I 35131 Padova, Italy\\
Istituto Nazionale di Fisica Nucleare, Sezione di Padova, \\
Via Marzolo 8, I 35131 Padova, Italy \\
Istituto Nazionale per la Fisica della Materia, Unit\'a di Milano, \\
Via Celoria 16, 20133 Milano, Italy

\end{center}

\vskip 0.8 truecm
\begin{center}
{\bf Abstract}
\end{center}
\vskip 0.5 truecm
\par
As pointed out by the authors of the comment quant-ph/9712046, 
in our paper quant-ph/9712030 we studied in detail the {\it metastability} 
of a Bose--Einstein Condensate (BEC) confined in an harmonic trap with 
zero--range interaction. As well known, the BEC with attractive zero--range 
interaction is not {\it stable} but can be {\it metastable}. 
In our paper we analyzed the role of dimensionality for 
the {\it metastability} of the BEC 
with attractive and repulsive interaction. 


\newpage

\par 
In a recent paper$^{1)}$ we studied the ground--state 
stability of a Bose--Einstein condensate (BEC) confined 
in an harmonic trap with repulsive and attractive zero--range interaction 
by minimizing the energy functional of the system$^{3)}$ given by
\beq
{E\over N} = \int d^3{\bf r} \; {\hbar^2\over 2m} |\nabla \Psi ({\bf r})|^2 
+ V_0({\bf r}) |\Psi ({\bf r})|^2 +{BN\over 2} |\Psi ({\bf r})|^4 \; ,
\eeq 
where $\Psi ({\bf r})$ is the wavefunction of the condensate, 
$V_0({\bf r})=m\omega^2 {\bf r}^2/2$ is the external potential 
of the trap, and $B={4\pi \hbar^2 a_s/m}$ 
is the scattering amplitude ($a_s$ is the s--wave scattering length). 
Obviously $N$ is the number of Bosons of the condensate. 
\par
In their comment$^{2)}$ Brosens, Devreese and Lemmens 
demonstrate the well-known result (implicit in our paper) 
that the BEC is unstable if there is more than 1 Boson. 
\par
It is important to stress that in our paper$^{1)}$ we studied not only 
the {\it stability} but also the {\it metastability} of the BEC 
by varying the spatial dimension of the system. 
In the case of repulsive interaction the system is stable and 
the BEC mean radius grows by increasing the number of Bosons. 
In the case of attractive interaction there is a metastable regime and 
the BEC mean radius decreases by increasing 
the number of Bosons: to zero if the system is one--dimensional and 
to a minimum radius, with a maximum number of Bosons, 
if the system is three--dimensional. 
\par
In the second part of their comment, the authors of Ref. 2 
claim that replacing the zero-range interaction 
by a short-range attractive interaction lifts the instability, and leads to 
a pronounced clustering, by which the particles leak out of the condensate. 
This is an interesting result and probably correct. 

\section*{Acknowledgments}
\par
The author is grateful to Prof. A. Parola and Prof. L. Reatto 
for stimulating discussions. 

\newpage

\section*{References}

\begin{description}

\item{\ 1.} L. Salasnich, quant-ph/9712030, to be published 
in Mod. Phys. Lett. B (98). 

\item{\ 2.} F. Brosens, J.T. Devreese and L.F. Lemmens, quant-ph/9712046.  

\item{\ 3.} E.P. Gross, Nuovo Cimento {\bf 20}, 454 (1961); 
L.P. Pitaevskii, Sov. Phys. JETP {\bf 13}, 451 (1961).

\end{description}

\end{document}